# Paid and hypothetical time preferences are the same:
# Lab, field and online evidence[1]

Pablo Brañas-Garza[†], Diego Jorrat[†], Antonio M. Espín[‡], and Angel Sánchez[§][♣]

[†]*LoyolaBehLab*, Universidad Loyola Andalucía, [‡]Department of Social Anthropology, Universidad de Granada, [§]GISC & IBiDat, Universidad Carlos III de Madrid

[♣]BIFI, Universidad de Zaragoza

September 19, 2020

**Abstract**

The use of hypothetical instead of real decision-making incentives remains under debate after decades of economic experiments. Standard incentivized experiments involve substantial monetary costs due to participants' earnings and often logistic costs as well. In time preferences experiments, which involve future payments, real payments are particularly problematic. Since immediate rewards frequently have lower transaction costs than delayed rewards in experimental tasks, among other issues, (quasi)hyperbolic functional forms cannot be accurately estimated. What if hypothetical payments provide accurate data which, moreover, avoid transaction cost problems? In this paper, we test whether the use of hypothetical - versus real - payments affects the elicitation of short-term and long-term discounting in a standard multiple price list task. One-out-of-ten participants probabilistic payment schemes are also considered. We analyze data from three studies: a lab experiment in Spain, a well-powered field experiment in Nigeria, and an online extension focused on probabilistic payments. Our results indicate that *paid and hypothetical time preferences are mostly the same* and, therefore, that hypothetical rewards are a good alternative to real rewards. However, our data suggest that probabilistic payments are not.

[1] We thank Giussepe Attanasi, Michele Belot, Klarita Gërxhan, Nagore Iriberri, Rosemarie Nagel, Pedro Rey-Biel, Amparo Urbano and audience at EUI-Florence, ESA2019, ESA2020, ECREEW-Seville, SAE-Alicante for their comments and suggestions. We also thank A. Amorós, J. F. Ferrero, E. Mesa and A. Nuñez who conducted the lab experiment; Laura Costica, Edwin Daniel (Hanovia Ltd) for data collection in Nigeria. Nigeria data where gathered during the pilots of a World Bank project (PIs: Ericka Rascon and Victor Orozco). Funding provided by the Ministry of Spain PGC2018-093506-B-I00 and PGC2018-098186-B- I00 and by grants PGC2018-098186-B-I00 (BASIC, FEDER/MICINN- AEI), PRACTICO-CM (Comunidad de Madrid), and CAVTIONS-CM-UC3M (Comunidad de Madrid/Universidad Carlos III de Madrid).

# 1 Introduction

Patience is becoming a major topic in Economics[2]. Patience refers to the preference for larger rewards in the future over smaller rewards in the present. Time discounting (TD) represents the loss of utility associated to any reward which is not obtained immediately but deferred in time. Patient individuals, therefore, display lower TD than impatient individuals.

Time preferences are relevant in a number of fields. *Health behavior*: There is evidence suggesting that experimental measures of individuals' patience negatively correlate with alcohol consumption, smoking behavior, and body mass index (Borghans and Golsteyn, 2006; Chabris et al., 2008; Sutter et al., 2013). *Education*: There is evidence that suggests that subjects with a high level of patience, i.e. lower TD, have better education outcomes (Golsteyn et al., 2014; Kirby et al., 2005; Duckworth and Seligman, 2005; Non and Tempelaar, 2016; Paola and Gioia, 2014), and are less likely to receive disciplinary referrals in school (Castillo et al., 2011) and to drop out from high school and college (Cadena and Keys, 2015). *Finance:* Patience is correlated with income, savings, credit card borrowing (negatively) and is a good predictor of the real-life wealth distribution (Tanaka et al., 2010; Giné et al., 2017; Ashraf et al., 2006; Meier and Sprenger, 2010, 2013; Epper et al., 2020). *Other domains:* Patience is also correlated with *cognitive ability* (Frederick, 2005; Dohmen et al., 2010; Bosch-Domènech et al., 2014), *criminal behavior* (Äkerlund et al., 2016), *divorce probability* (Paola and Gioia, 2017), and *social behavior* (Dewitte and Cremer, 2001; Rachlin, 2002; Espín et al., 2012; Espín et al., 2015).

Beyond these individual-level relationships, there is new macro-level evidence connecting patience with economic development: patience seems to be a key determinant of GDP per capita, and its effect seems to arise through physical and human capital accumulation and productivity (Dohmen et al., 2018). Other macro-level relationships include intergroup discrimination (negatively), life expectancy (positively) and infant mortality (negatively) (Bulley and Pepper, 2017; Espín et al., 2019b). Recently, there is also an interest in studying the effect of in-the-classroom interventions on the consistency of inter-temporal decisions (Alan and Ertac, 2018; Lührmann et al., 2018), and the causal relationship between education and patience (Perez-Arce, 2017; Kim et al., 2018).

---

[2] According to the Scopus database, the amount of papers containing "time preference" in the title is about four times greater in 2018 compared to 2000.

With some exceptions, TD is typically elicited using Multiple Price Lists (MPL) tasks, originally designed by Coller and Williams (1999). In these tasks, subjects decide whether to take ("sooner" option) or save ("later" option) a certain amount of money over a series of independent choices with increasing interest rates.[3]

Although for experimental economists the use of monetary incentives is a must, some of the most acclaimed papers on TD – for instance Kirby et al. (2005); Ashraf et al. (2006); Golsteyn et al. (2014); Cadena and Keys (2015); Dohmen et al. (2018) just to name a few – do not pay participants real monetary incentives, that is, decisions are hypothetical. But, are choices over hypothetical rewards equally informative as incentivized time preferences?

Besides the obvious monetary and logistic costs, the use of monetary incentives in TD elicitation tasks is quite controversial and challenging. First, transaction costs and payment reliability need to be constant across options regardless of the payment date (Cohen et al., 2020). For example, future payments must be just as reliable as immediate payments: if a subject feels that the later reward might be not delivered then s/he will be more willing to choose the sooner reward to avoid the uncertainty, not due to TD. This problem is more prominent in the field, especially in low-income populations, where many people are unbanked or change their cell-phone numbers (often used to contact them for future payments) frequently. All these factors increase the uncertainty associated to future payments and, consequently, the probability that a subject will prefer the sooner option for reasons other than time preferences. This may compromise the estimation of individual discount rates using one of the most common functional forms, i.e.

---

[3] There is another experimental method commonly used to elicit time preferences, the so-called Convex Time Budgets (CTB) designed by Andreoni and Sprenger (2012). Some recent experimental research in the field (Meier and Sprenger, 2013; Giné et al., 2017; Lührmann et al., 2018) implements this method.

(quasi)hyperbolic preferences (Laibson, 1997), given that immediate vs. delayed payoffs need to be considered in the task.

Second, subjects' choices only reflect their time preference if they are liquidity constrained and have the impossibility to make inter-temporal arbitrage (Frederick et al., 2002; Lührmann et al., 2018). Otherwise, subjects should compare the interest rate offered in the task (r) with the market interest rate ($\bar{r}$). For $r < \bar{r}$ subjects can just take the money today and bring it to a bank deposit. For $r > \bar{r}$, they should save all in the task. Therefore, the task will elicit estimated discount rates which are contaminated and do not reflect pure time preference. Third, a higher expectation of future inflation may lead an individual to prefer sooner-smaller rewards without the influence of time preference, simply because the money is worthless in the future (Frederick et al., 2002; Martín et al., 2019). This is a critical problem in volatile economies, where people typically face high inflation rates. Fourth, there is a serious problem with data privacy. If subjects need to be paid by bank transfer, then they need to release private information (e.g. phone number, bank number). Relatedly, there is a growing interest on the development, malleability and stability of time preferences (Alan and Ertac, 2018; Lührmann et al., 2018; Perez-Arce, 2017; Kim et al., 2018). To study these questions, researchers usually need to employ adolescent or children samples. In these cases, using real money requires especial parental consent which usually makes more difficult to obtain approval from Ethics Committees.

Thus, if the use of real incentives in TD elicitation is expensive, may induce biased estimates and is, in general, so problematic, why we pay decisions at all? This question is especially important in field experiments. Given the potential benefits of using hypothetical rewards, it is fair to say that the topic remains relatively understudied.

Previous evidence directly comparing both mechanisms (real vs. hypothetical), is typically based on lab experiments with small samples and low power. Using a within-subject design, Madden et al. (2003) compared the hyperbolic discount rates estimated from students' choices over incentivized and hypothetical rewards and found no differences between both measurements. Johnson and Bickel (2002), also using a within-subject design, found no significant differences between incentivized and hypothetical choices. Madden et al. (2004) replicated the (null) results using a between-subject design. Bickel et al. (2009) compared TD choices with real and hypothetical money using neuroimaging and found no significant behavioral or neurobiological differences. Lawyer et al. (2011), using a between-subject design with non-student population, compared hypothetical and incentivized choices and found no significant differences in either nicotine-dependent or non-dependent samples. Last, Matusiewicz et al. (2013), tested the same hypothesis in the lab and found no significant differences between incentivized and hypothetical choices, neither in an initial measurement nor in a retest one week after. On the contrary, Coller and Williams (1999) found a weak significant difference between both mechanisms, but the authors assigned treatment status at the session rather than individual level, leading to lack-of-balance problems.

In sum, these papers suffer from several issues. They are either confined to lab experiments and student subjects, or their sample sizes are rather small (between 6 and 60 subjects), or do not perform a correct random assignment of subjects into treatments, as is the case of Coller and Willians (1999).

Evidence from field experiments is even scarcer. Harrison et al. (2002), in a field experiment in Denmark, examined if variation in the between-subjects random incentive system had an impact in TD. Keeping the number of paid subjects per session fixed but increasing the session size (hence the probability of being paid decreases) they found no significant difference. Ubfal (2016) estimated discount rates for six different goods among rural Ugandan households using hypothetical and incentivized choices and found no significant differences between the two. However, in Ubfal (2016) the within-subjects design employed was not counterbalanced; all individuals decided in the same order. In both studies, subjects were not randomly assigned to treatments individually since studying the use of hypothetical vs. real incentives was not the studies' main research goal.

Apart from these papers studying level differences between both payment methods, there is also evidence that choices in hypothetical tasks are correlated with choices in incentivized tasks, and predict similar behaviors. Using a sample of 409 German undergraduates, Falk et al. (2015) validated a hypothetical TD task using a real-incentives task. They found that both measures were highly correlated and predicted out-of-sample behavior. However, they did not analyze whether the use of hypothetical rewards generates a systematic bias in elicited discount rates.

This paper aims to fill this knowledge gap and quantify the impact of hypothetical vs. real rewards on TD elicitation. We study this in a comprehensive manner, across different populations and settings. We first analyze data from a lab experiment in Spain and a lab-in-the-field experiment in rural Nigeria. In addition, in both samples we examine the impact of another commonly used payment method, consisting of paying only a fraction of subjects (in our case, 10%). The latter analysis is complemented with data from an online experiment. In all cases, subjects are randomly assigned to a different payment mechanism and the data allow us to test for effects on both short- and long-term discounting decisions as well as on the individual parameters obtained assuming quasi-hyperbolic preferences (i.e. the beta and delta discount factors).

Even if the studies above suffered from several problems, the main message so far seems to be that hypothetical incentives do not induce estimation bias. However, without a more systematic and robust analysis, such a conclusion would be too premature.

Our results are rather clear and help close the debate: choices over hypothetical rewards do not differ from choices over real rewards in TD tasks. In fact, we believe that the accumulated evidence now allows us to conclude that hypothetical rewards do not induce estimation bias. On the other hand, we find that probabilistic payment schemes make a difference and, therefore, do not appear to be the best alternative to real incentives. Thus, while we believe that the debate regarding the use of hypothetical rewards might be considered as closed with our results, the debate regarding the use of probabilistic payment remains absolutely open[4]. For example, it is unknown whether different probabilities of being paid have different effects on decisions.

---

[4] Cubitt et al. (1998) and Brañas-Garza et al. (2020) study BRIS payoffs in risk preferences. Both papers found a significant difference with respect to real payments, with lower risk aversion in the BRIS treatment.

These results are more than a purely methodological contribution. They have important implications for the budgets of research groups running TD experiments, especially in the field in developing and volatile economies, as well as for the design of large-scale representative surveys which increasingly include behavioral tasks to estimate economic preferences in the population. Of course, these results cannot be extrapolated to the use of hypothetical rewards in measures of risk or social preferences and there are indeed several studies showing that, quite clearly in the latter case, real money matters (e.g. Clot et al., 2018; Holt and Laury, 2002).

In the online experiment, we also assess the robustness of hypothetical payoffs to different design features which are typical in experiments and surveys, such as changing the order between the short-term and long-term blocks of TD decisions, having subjects playing other tasks (games) before the TD elicitation, and whether those games are also hypothetical or not. We find that playing other games first, regardless of whether these are incentivized, leads to increased patience, especially for short-term discounting. However, while in principle playing the TD first is expected to yield a cleaner measurement of patience, our results do not necessarily imply that the TD elicited after other games is biased – it might be argued, for example, that having played other games reduces the anxiety associated to facing the TD task for the first time so that the increased patience observed is actually closer to the true value. This is an interesting avenue for future research.

The rest of the paper is structured in seven sections. The second section focuses on the main research questions, whereas the third reports on general features of the design. Sections four and five analyze the lab and the field experiment, respectively. A robustness check of the results in the lab and field is conducted along the sixth. Section seven focuses on the online experiment. The last section discusses the results and the contribution of the paper.

## 2 Questions to be addressed

The main research question of this paper is:

*Q1: Do hypothetical payments (H) provide the same outcome as real payments (R) in TD elicitation tasks?*

Recently, a number of studies have used an intermediate solution as a cost-saving alternative to real payments: paying one subject out of X, rather than all of them. Very often, X is set to ten. This means that participants have 10% probability of getting paid for real. In the so-called Between-subjects Random Incentivized System (BRIS), only a fraction of individuals is randomly selected to receive the real payments, and participants are aware of this probability ex-ante (Baltussen, 2012). The associated monetary and logistic costs decrease proportionally but the rest of the problems (lack of trust, inflation, privacy and financial arbitrage) remain untouched. As a second goal of this research, we will compare this probabilistic mechanism with real payments (i.e. 10% vs. 100% probability). We rephrase Q1 as follows to account for BRIS:

*Q2: Do one-out-of-ten payments (B) provide the same outcome as real payments (R) in TD elicitation tasks?*

To address Q1 and Q2, we conduct two experiments. First, we run a lab experiment with university students in Seville (Spain) and a lab-in-the-field experiment in the Kano province in Nigeria.

Besides Q1 and Q2, we analyze data from an online experiment that can offer several insights. Since in the online sample there are no Real payments, we study Q2 indirectly: whether BRIS and Hypothetical schemes are the same. We also check this parallel question in the lab and field studies by testing the difference between the estimates for *H* and *B*.

Finally, the online data allow us to test whether hypothetical payoffs are sensitive to different design features, some of which are typical in large-scale experiments and surveys: *Within-task Order*, that is, whether short- or long-term TD decisions are made first; *Position of the task*, that is, whether other games are played before the TD elicitation; and *Previous paid tasks*, that is, whether the games preceding the TD task use the hypothetical or the BRIS mechanism.

## 3 Treatments, balance and MPL task

In the three studies, we follow the same protocol: participants are randomly assigned to one of the treatment arms (H/R/B in the lab and field experiments, H/B in the online experiment).

The randomization allows us to evaluate the *causal impact* of different payment schemes over the estimated TD. Throughout this section we explain the randomization, the resulting samples, and the MPL task.

### 3.1 Treatments

We compare 3 treatments that differ in the probability of being paid (from 1 to 0):

- **R**: Earnings with probability $p = 1$, where all subjects get a real payment.

- **B**: Earnings with probability $p = 1/10$, where 1 subject out of 10 get a real payment.

- **H**: Earnings with probability $p = 0$, where none of the subjects get a real payment.

All the subjects were informed of their payment scheme ex-ante but they were not aware of the existence of other payment schemes, i.e. treatments. It should be noted that in the online experiment (Study III) we only ran treatments B and H.

### 3.2 Sample and balance across experiments

Table 1 shows the balance of the randomization across treatments in each experiment. In the lab experiment (top panel), we can see that all the individual characteristics (age, gender and score on the CRT) are balanced between the treatments, except for one marginally significant difference in age: in the BRIS treatment (B), individuals were on average 1.2 years younger than in the R treatment ($p = 0.064$).

In the field experiment (central panel), we observe significant differences in age, as participants in the H treatment were 2.4 years older than those in the R treatment ($p = 0.024$). It should be noted that risk preferences were measured only for half of the sample ($n = 360$). Still, the treatments were balanced in risk preferences as well.

The bottom panel of Table 1 displays the between-treatments balance for the online experiment. As mentioned earlier, the design of the online experiment is different. In this study we only compare treatments B and H. We observe marginally significant differences in terms of age and female proportion: the participants in the H treatment were 1.6 years older than those in the B treatment ($p = 0.081$) and the fraction of females was 6% higher ($p = 0.088$).

Table 1: **Balance across treatments in Studies I, II and III**

|  | obs. | $mean_R$ | $H - R$ | p | $B - R$ | P |
|---|---|---|---|---|---|---|
| *Study I: Lab* | | | | | | |
| Age | 119 | 21.846 | -0.471 | 0.463 | -1.196 | 0.064* |
| Female | 119 | 0.385 | 0.065 | 0.563 | 0.115 | 0.308 |
| CRT | 118 | 1.282 | -0.332 | 0.153 | -0.308 | 0.188 |
| *Study II: Field* | | | | | | |
| Age | 721 | 39.238 | 2.421 | 0.024** | -0.442 | 0.678 |
| Female | 721 | 0.527 | 0.049 | 0.283 | 0.030 | 0.512 |
| Education | 721 | 7.715 | 0.211 | 0.714 | 0.064 | 0.915 |
| Sufficient[+] | 721 | 0.787 | 0.006 | 0.870 | 0.027 | 0.464 |
| Risky choices | 360 | 1.917 | 0.004 | 0.983 | -0.226 | 0.252 |
| | | $mean_B$ | $H - B$ | p | | |
| *Study III: Online* | | | | | | |
| Age | 632 | 38.511 | 1.601 | 0.081* | | |
| Female | 637 | 0.238 | 0.060 | 0.088* | | |
| Education | 637 | 6.581 | 0.003 | 0.983 | | |
| Household income | 635 | 1,031.512 | -22.114 | 0.597 | | |
| Risky choices | 637 | 0.463 | -0.060 | 0.233 | | |

Note: Inference was made using OLS regression with robust standard errors. *** $p < 0.01$, ** $p < 0.05$, * $p < 0.1$. [+] Sufficient refers to self-reporting having enough money to feed the family. *R* refers to Real, *H* to Hypothetical and *B* to BRIS.

However, none of the four significant or marginally significant differences between treatments would survive a Bonferroni-like correction for multiple testing. Thus, the results in Table 1 suggest that the randomization worked properly. Having balance between treatments is essential to isolate the impact of different payments mechanisms on TD choices. The results imply that our sub-samples are nearly identical and therefore we can test *causal effects of incentives* on decisions. Moreover, the regression analysis will allow us to control for potential confounds.

**3.3 Eliciting time preferences**

Our instrument to measure time preferences was adapted from Coller and Williams (1999) and Espín et al. (2012). Similar tasks have been used for instance in Burks et al. (2012), Espín et al. (2015), (2019a), and Martín et al. (2019) – see Frederick et al. (2002) for an extensive survey. Participants made a total of 20 binary choices between a sooner smaller amount of money and a later but larger amount in two blocks of ten decisions each. The first block involves choosing between a no-delay option ("today") and a one-month delay option, while the second block involves a one-month delay option and a seven-month delay option. We refer to these two blocks as short-term and long-term TD decisions, respectively. We used the same amounts in both blocks, whereas interest rates vary according to the time horizon considered in each block. The amount of the sooner payoff was fixed across decisions and the amount of the later payoff increased in interest rate from decision 1 to decision 10 (see Table 2). This method is known in the literature as Multiple Price List (MPL).

The protocol described above allows us to compute the beta and delta parameters ($\beta_i$, $\delta_i$) of a quasi-hyperbolic discount function (Burks et al., 2012; Laibson, 1997; McClure et al., 2004; Phelps and Pollak, 1968). The beta-delta model formalizes the individual's discount function as $V_d=\beta\delta^t V_u$, where $V_d$ is the discounted psychological value of a reward with (undiscounted) value $V_u$ which will be received in $t$ time units. $\beta$ and $\delta$ $\varepsilon$ (0, 1] are the "beta" and "delta" discount factors, respectively. The higher these discount factors the more patient the individual is, as delayed rewards are valued more (i.e. they are discounted less). The beta discount factor refers to present bias, that is, the value of any non-immediate reward is discounted by a fixed proportion $\beta$, regardless of the delay. The delta discount factor captures "long-term discounting" in an exponential functional form, that is, for each unit of time that constitutes the delay to delivery, the value of a reward is discounted by $\delta$. This model thus allows for a possible difference between short-term and long-term discounting, and has been shown to predict outcomes better than other formulations (Burks et al., 2012).

We opted for the non-delayed option ("today") because we wanted to test whether there are differences in beta, i.e. present bias, between *H* and *R*. Present bias refers to the apparent tendency of (some) individuals to assign a premium to immediate rewards (McClure et al. 2004; Takeuchi, 2011). It is reasonable to expect that the "today" option induces stronger differences between hypothetical and real rewards because the immediacy premium might partly capture differences in uncertainty or transaction costs between immediate and non-immediate rewards (Chabris et al., 2008), which are absent in hypothetical scenarios. There is evidence that delaying the sooner option by one day helps to avoid possible confounds such as differential transaction costs between payment dates or trust issues (Sozou, 1998). However, without a truly immediate option, the beta parameter cannot be accurately estimated. In our design, with a "today" option, we therefore expected to find the strongest differences between the *H* and *R* treatments in present bias $\beta$, or short-term discounting.

Table 2: **MPLs design across experiments**

| Lab (Euros) | | Field (Nairas) | | Online (Euros) | | Monthly interest rate | |
|---|---|---|---|---|---|---|---|
| Sooner | Later | Sooner | Later | Sooner | Later | Short | Long |
| 10 | 10 | 400 | 400 | 30 | 30 | 0.00% | 0.00% |
| 10 | 10.7 | 400 | 427 | 30 | 32 | 6.70% | 1.12% |
| 10 | 11.3 | 400 | 453 | 30 | 34 | 13.40% | 2.23% |
| 10 | 12 | 400 | 480 | 30 | 36 | 20.10% | 3.35% |
| 10 | 12.7 | 400 | 507 | 30 | 38 | 26.70% | 4.45% |
| 10 | 13.3 | 400 | 533 | 30 | 40 | 33.30% | 5.55% |
| 10 | 14 | 400 | 560 | 30 | 42 | 40.00% | 6.67% |
| 10 | 14.7 | 400 | 587 | 30 | 44 | 46.70% | 7.78% |
| 10 | 15.3 | 400 | 613 | 30 | 46 | 53.40% | 8.90% |
| 10 | 16 | 400 | 640 | 30 | 48 | 60.00% | 10.00% |

Notes: Monthly simple interest rates are displayed. The interest rates differ between the short-term and long-term blocks because the delays considered are one month and six months, respectively.

In each block, we obtained the switching point where a participant was indifferent between both options. Following the protocol introduced by Burks et al. (2012), we computed the $\beta$ and $\delta$ for each participant. The time units were defined in months. As standard, we assume that utility is linear over the relevant range.

Table 3 provides the descriptive statistics for the short-term ($\beta$) and long-term ($\delta$) discount factors obtained in each study. Note that, following standard methodology, we do not restrict $\beta$ to be smaller than one, so that some individuals might display "future bias" (Takeuchi, 2011; Jackson and Yariv, 2014; Balakrishnan et al., 2020).

The proportion of inconsistent choices (multiple switching or non-monotonic patterns) differs slightly across studies and blocks of the task. In the lab experiment, 3% of participants made inconsistent choices in the short-term block (today vs. one-month), whereas none did so in the long-term block (one-month vs. seven-month). In the field experiment, only 1% of the participants made inconsistent choices in either of the two blocks. It should be noted that the field experiment was conducted by enumerators who were trained to avoid inconsistencies. In the online experiment, 2% of the subjects made inconsistent choices in the short-term block, 2.5% in the long-term block.

Table 3: **Discount factors and number of later allocations by study**

| Variable | Mean | sd | min | max | incons(%). |
|---|---|---|---|---|---|
| *Study I: Lab* | | | | | |
| Beta | 0.827 | 0.097 | 0.652 | 1.049 | 2.80% |
| Delta | 0.937 | 0.02 | 0.918 | 0.99 | 0.00% |
| # later alloc. (short) | 5.233 | 2.502 | 0 | 10 | - |
| # later alloc. (long) | 2.617 | 2.577 | 0 | 9 | - |
| | | | | | |
| *Study II: Field* | | | | | |
| Beta | 0.744 | 0.142 | 0.612 | 1.088 | 0.80% |
| Delta | 0.931 | 0.027 | 0.918 | 1.000 | 0.70% |
| # later alloc. (short) | 2.617 | 3.898 | 0 | 10 | - |
| # later alloc. (long) | 1.648 | 3.390 | 0 | 10 | - |
| | | | | | |
| *Study III: Online* | | | | | |
| Beta | 0.888 | 0.11 | 0.606 | 1.088 | 2.10% |
| Delta | 0.955 | 0.028 | 0.918 | 1.000 | 2.50% |
| # later alloc. (short) | 6.937 | 2.940 | 0 | 10 | - |
| # later alloc. (long) | 4.871 | 3.506 | 0 | 10 | - |

For the sake of completeness, in the empirical analysis we also consider the number of later allocations in the short- and long-term blocks as an alternative measure of individuals' patience. Unlike beta and delta, these measures are not parameterized and consider individuals making both consistent and inconsistent decisions ($\beta$ and $\delta$ cannot be computed for inconsistent individuals). Figure 1 shows the distribution of the number of later allocations by study.

Figure 1: **Short- and long-term later allocations by study**

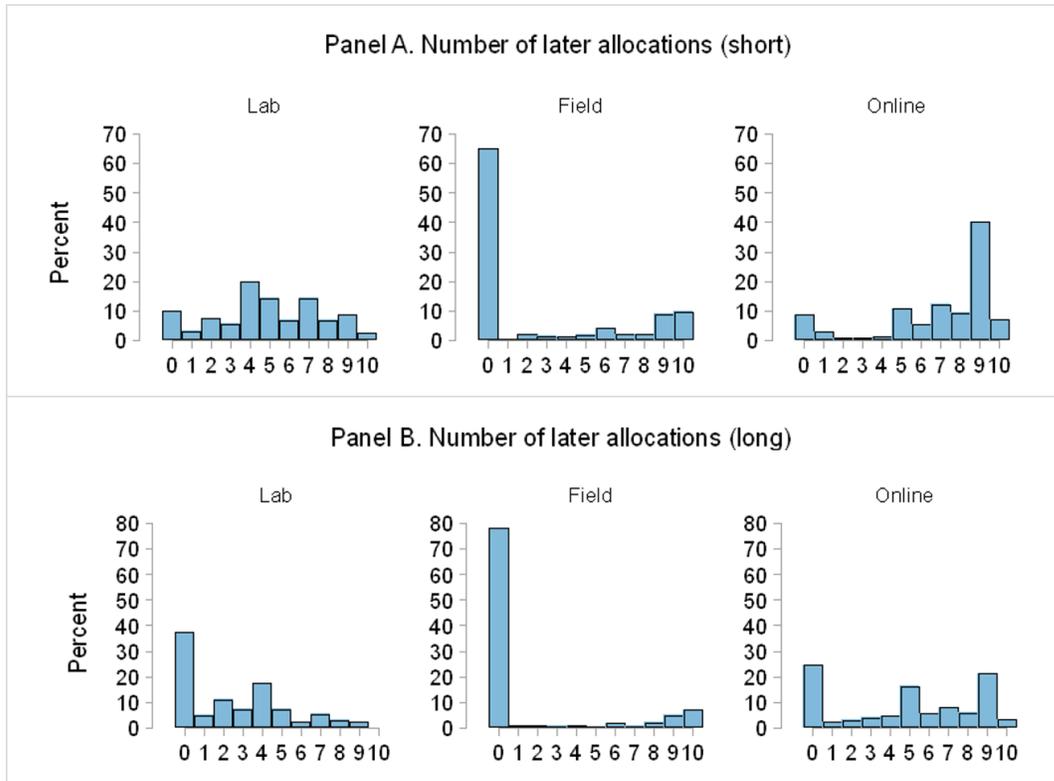

In Nigeria (field), 65% and 80% of the subjects chose the sooner option in all the ten decisions (i.e. number of later allocations = 0) on the short-term and long-term block, respectively. Such a high percentage of people choosing always the sooner reward is common in the literature, particularly in field experiments: Martín et al. (2019) found that 48% of Spanish Gitanos choose the sooner option in all decisions in a MPL with 20 decisions (with delay identical to our long-term block, i.e. one month vs. seven months). Of course, increasing the interest rate associated to the delayed reward would have likely reduced the number of sooner allocations in the field study. Yet this would have been counterproductive in the two experiments conducted with Spanish participants, i.e. the lab and the online experiment, where the data are quite well distributed. Our procedure provides comparable decisions, and therefore comparable estimates, across samples. Note that the Global Preferences Survey developed by Falk et al. (2018) ranks Nigeria in 49th position (out of 76 countries) in terms of patience, whereas Spain occupies the 18th position. In fact, the percentage of all-sooner choices in either block are much smaller in the lab and online experiments. Regardless of the time-horizon considered, hence, participants in Nigeria appear to be more impatient than Spanish participants. However, the goal of this study is not to compare patience between Spanish and Nigerian participants.

## 4 Study I: The lab experiment

Compared to the field, the lab provides for a more controlled test of whether different reward schemes affect TD measures. In the lab, experimenters have a higher degree of control over the environment and can ensure greater credibility for future payments. The lab typically also has some drawbacks though: participants are university students, self-selected into the experiment and with a relatively high socioeconomic status.

### 4.1 Implementation and sample

We ran the lab experiment in the University of Seville and the Pablo de Olavide University, both in Seville, Spain, between April and May 2019. Participants were recruited in the two campuses using flyers and the School of Economics website. Among the 473 subjects who signed up, 120 were randomly assigned to this study. Then they were randomly assigned to treatments *R*, *H* or *B* with probability 1/3. One participant had to leave a few minutes after the experiment started.

The sample is composed of students from Business (31%), Law Economics (24%), Marketing (20%), Economics (16%), and other degrees. The average participant age was 22 and 39% were female.

For final payments, one out of the 20 decisions was randomly selected. In the R condition, all participants were paid the amount associated to their choice in that decision at the corresponding date (either "today", or in one month, or in seven months), whereas in the *B* treatment we randomly selected 10% of them to receive the money. No participant was selected for payment in the *H* treatment.

We offered participants the possibility of bank transfers, but only about 40% selected this option. The remaining 60% was paid in cash at the University the day associated to the randomly selected decision. All participants received a show-up fee of 4 euros.

### 4.3 Ethics

All participants were informed about the content of the experiment before participation. Participants signed an informed consent. The study was approved by the Ethics Committee of Loyola Andalucía University.

### 4.4 Results

In Table 4 we show the impact of hypothetical ($H$) and BRIS ($B$), versus real ($R$), incentives on individual's patience using OLS regressions with different specifications. Columns 1 to 4 display the results when the dependent variable is $\beta$ or $\delta$ from the beta-delta model. In columns 5 to 8 the dependent variable is the number of later allocations in the two blocks (short-term or long-term). The regressions in columns 3, 4, 7, and 8 control for age, gender, and CRT score. We use CRT as a proxy of participants cognitive abilities. None of these variables are significant ($p>0.2$).

#### 4.4.1 Are hypothetical and real choices different (Q1)?

We study first whether using hypothetical rewards or paying for real yield different choices. Columns 1 and 2 show that the $H$ dummy has no significant impact on beta or delta ($p > 0.89$), suggesting that hypothetical decisions do not differ from real incentivized decisions ($R$). After adding the control variables (columns 3 and 4), the coefficients remain non-significant ($p > 0.87$). Regarding the number of later allocations (columns 5-8), $H$ does not yield significant estimates on either the short-term or the long-term block (p > 0.86 without controls, p > 0.78 with controls).

One particular concern regarding the use of hypothetical decisions is that they might increase variance, i.e. noisy decision making (for example, due to lack of attention), but not necessarily change the average response (Camerer and Hogarth, 1999). Our lab results indicate that indeed the average response does not differ between $H$ and $R$. To study differences in variance, a summary of the results of a series of variance ratio tests is included in Table 5. Since the hypothesis is that real incentives trigger less noisy decisions, we conduct one-tailed tests against this hypothesis. Panel *i)* shows the standard deviation of the mean for each variable by treatment. It can be seen that, against our hypothesis, *R* yields the highest SD. Panel *ii)* confirms that the ratio of the standard deviation between *R* and *H* is not significantly lower than one for any of the outcome variables (p > 0.80).

Table 4: **Estimated differences between treatments (Study I)**

|  | (1) beta | (2) delta | (3) beta | (4) delta | (5) #later alloc. (short) | (6) #later alloc. (long) | (7) #later alloc. (short) | (8) #later alloc. (long) |
| --- | --- | --- | --- | --- | --- | --- | --- | --- |
| H | -0.003 | 0.000 | 0.001 | 0.001 | -0.000 | 0.100 | 0.106 | 0.171 |
|  | (0.022) | (0.004) | (0.023) | (0.005) | (0.547) | (0.574) | (0.579) | (0.612) |
|  | [0.899] | [0.943] | [0.968] | [0.872] | [1.000] | [0.862] | [0.855] | [0.780] |
| B | -0.037 | -0.003 | -0.030 | -0.002 | -1.100* | -0.350 | -0.897 | -0.274 |
|  | (0.023) | (0.005) | (0.025) | (0.005) | (0.579) | (0.601) | (0.637) | (0.646) |
|  | [0.102] | [0.571] | [0.236] | [0.656] | [0.060] | [0.561] | [0.162] | [0.672] |
| Constant | 0.840*** | 0.938*** | 0.835*** | 0.945*** | 5.600*** | 2.700*** | 5.389*** | 3.470* |
|  | (0.016) | (0.003) | (0.060) | (0.013) | (0.422) | (0.433) | (1.471) | (1.800) |
|  | [0.000] | [0.000] | [0.000] | [0.000] | [0.000] | [0.000] | [0.000] | [0.056] |
| Observations | 116 | 120 | 114 | 118 | 120 | 120 | 118 | 118 |
| R-squared | 0.030 | 0.005 | 0.063 | 0.015 | 0.043 | 0.006 | 0.066 | 0.015 |
| Controls | No | No | Yes | Yes | No | No | Yes | Yes |
| MCG+ | 0.839 | 0.937 | 0.839 | 0.937 | 5.601 | 2.701 | 5.601 | 2.701 |

Note: OLS estimates. Robust standard errors in parentheses and p-values in brackets. ***p < 0.01, **p < 0.05, *p < 0.1. Controls are age, gender, and CRT score. + MCG refers to the Mean for the Control Group (R treatment). Subjects making inconsistent choices are excluded from the analysis of beta-delta.

Taking into account all the evidence, we can summarize the main result from Study I as follows:

**R1 (lab: H vs. R)**: *Using hypothetical vs. real payoffs does not generate different discount factors (β-δ model) or different numbers of later allocations, neither in terms of averages nor in terms of variance of responses.*

Table 5: **Variance ratio test for the outcome variables (Study I)**

|  | (1) Beta | (2) Delta | (3) #later alloc. (short) | (4) #later alloc. (long) |
|---|---|---|---|---|
| *i) Standard deviation by treatment* | | | | |
| $SD(R)$ | 0.103 | 0.021 | 2.668 | 2.738 |
| $SD(H)$ | 0.089 | 0.018 | 2.204 | 2.388 |
| $SD(B)$ | 0.098 | 0.020 | 2.511 | 2.636 |
| *ii) R vs H* | | | | |
| $SD(R)/SD(H)$ | 1.153 | 1.167 | 1.210 | 1.146 |
| $P\ (ratio < 1)$ | 0.812 | 0.841 | 0.881 | 0.802 |
| *iii) R vs B* | | | | |
| $SD(R)/SD(B)$ | 1.051 | 1.050 | 1.063 | 1.039 |
| $P\ (ratio < 1)$ | 0.622 | 0.552 | 0.6461 | 0.593 |

Note: The null hypothesis is that the ratio between the standard deviation of the variable in the R group and the standard deviation in the H (or B) group is smaller than 1.

### 4.4.2 Are BRIS and real choices different (Q2)?

Now we focus on the comparison between BRIS and real payments. Columns 1 and 2 of Table 4 show that the dummy for BRIS ($B$) payments does not have a significant impact on beta or delta ($p > 0.10$), suggesting that BRIS decisions do not differ from fully incentivized decisions ($R$). This result holds after adding controls (columns 3 and 4, $p > 0.23$). On the other hand, column 5 shows that $B$ yields a negative and marginally significant effect on the number of later allocations in the short-term block ($p = 0.06$). After adding the control variables (column 7), however, $B$ is no longer significant ($p = 0.16$). No effect is found for the number of later allocations in the long-term block (columns 6 and 8, $p > 0.56$).

Panel *iii)* in Table 5 shows that the ratio of the standard deviation between $R$ and $B$ is not significantly lower than one for any of the outcome variables ($p > 0.55$).

Considering all the evidence, we can summarize these results as follows:

**R2 (lab: B vs. R)**: *Paying one out of ten vs. all subjects does not generate different discount factors (β-δ model) or different number of later allocations, neither in terms of averages nor in terms of variance of responses. If anything, the number of later allocations in the short-term block may be weakly affected (with B < R).*

# 5. Study II: The field experiment

Providing a precise measurement of TD in the field is a major concern for interventions in developing countries. Since recent studies (Levitt et al., 2016; Giné et al., 2017) have shown that time preferences have a decisive impact on treatment effects in that more patient individuals are more affected – for instance in educational interventions –, hence it is crucial to have an accurate instrument to measure them. However, as mentioned, the use of standard incentivized tasks in the field is especially complicated.

Hypothetical elicitation not only eliminates monetary costs but also ameliorates the other issues. Because of these reasons, hypothetical tasks are becoming popular in behavioral development economics. Based on the (fair) expectation that paying some subjects will provide more accurate estimates than paying none, the BRIS device is being often used as a cost-saving alternative to real payments. However, without a systematic analysis of the implications of the BRIS method, such an expectation may be unjustified. In Study II, we explore the effect of hypothetical and BRIS methods in the field, with a more heterogenous sample compared to the lab. In addition, since the null results of Study I might be affected by low statistical power, Study II provides a much larger sample and therefore a more powerful analysis.

We ran a lab-in-the-field experiment in the Kano province, Northern Nigeria, in order to test Q1 and Q2.

## 5.1 Implementation and sample

The experiment was conducted in seven villages in the Kano province: Albasu, Daho, Farantama, and Panda in a first wave; and Dorayi, Ja'en, and Gidan Maharba in a second wave. The first wave was conducted in November 2018 and the second in April 2019.

721 households were randomly selected to obtain a representative sample of the study area with the eligibility criterion of having at least one child between 6 and 9 years old[5]. Each household in the total sample was randomly assigned to one of the three

---

[5] We followed this criterion because the experiment was part of a much larger intervention conducted by the DIME (The World Bank). In addition, the random selection of households followed a geographical criterion based on their distance from the catchment areas of the local schools.

treatments (*H/R/B*) with 1/3 probability.

As is standard in the field, the experiment was conducted by enumerators, which implies that the instructions were read – and often explained – by the enumerator. 62 enumerators were hired and trained for the fieldwork.

Enumerators were given a list of households to visit and a tablet to conduct the interviews. The random allocation of households to treatments was computerized and the enumerators did not have any influence on such selection. Enumerators conducted face-to-face interviews in the households and only one person was interviewed per household.

The resulting sample size was $n = 721$ (by treatments, *R*: 239, *B*: 246, *H*: 236). Subjects were fully aware of their payment scheme. 55% of the participants were female and the average age was 38 years old. Also, 56% had primary education, 28% had completed secondary education and 16% had tertiary education. 79% reported to have sufficient money to feed the family in the last week.

The experiment consisted of four tasks: coordination games, expectations, time discounting and risk preferences. The TD task was always placed third. The payment scheme was hold constant across the entire experiment. That is to say, participants in the *R* treatment performed all the four tasks with real money, whereas participants in the *H* treatment performed all the four tasks with hypothetical money. The same applies to the BRIS treatment.

To elicit time preferences, we used the very same MPL (and same interest rates) as in the lab experiment. Table 2 shows the payments. We re-calculated payments in order be able to pay about one-day average wage for the entire experiment (equal to 1080 Nairas, or 3 US$). This resulted in a minimum payment of 400 Nairas in the TD task.

We randomly selected one of the 20 MPL choices to pay the TD task. For all participants in the *R* treatment and for the randomly selected 10% in the *B* treatment, we made the payments charging their cell phones with the chosen amount at the date of the selected decision.

## 5.2 Ethics

The study was approved by the Ethics Committees of Middlesex University London and IRBSolutions (US). All participants signed an informed consent.

## 5.3 Results

Table 6 provides the main results of Study II. We follow the same regression analysis as in Table 4. All regressions control for enumerator fixed effects; and only columns 3, 4, 7, and 8 include controls for age, gender, education level (from 1 = "no education", to 19 = "postgraduate") and income (=1 if they report to have enough money to feed the family). None of these control variables are significant in the regressions ($p>0.18$) except *education* ($p=0.08$). Nonetheless, some enumerator dummies yield significance, implying that enumerators did have an influence on the outcomes and therefore regressions should control for this.

### 5.3.1. Are hypothetical and real choices different (Q1)?

As in the case of the lab experiment, we first test the main question of the paper: Do $H$ and $R$ yield different TD choices (Q1)?

Columns 1 to 2 in Table 6 show that the use of hypothetical payments ($H$) does not have any significant impact on beta or delta ($p = 0.78$ and $p = 0.19$, respectively). This result holds after adding the control variables (columns 3 and 4). Regarding the number of later allocations (columns 5-8), $H$ does not yield significant estimates on either the short-term or the long-term block ($p > 0.19$ in both cases).

All in all, Study II yields the same (null) results for $H$ vs. $R$ as Study I, hence R1 is replicated.

Regarding the variance of responses, Table 7 shows the variance ratio test for each outcome variable. It can be seen in panel *i)* that, except for beta, the $R$ treatment displays the lowest SD, as hypothesized. Yet, panel *ii)* shows that the difference between $R$ and $H$ is not significant for either beta or the number of later allocations in the short-term block ($p > 0.41$), while it is significant for both delta and the number of later allocations in the long-term block ($p < 0.01$).

Table 6: **Estimated differences between treatments (Study II)**

|   | (1) beta | (2) delta | (3) beta | (4) delta | (5) #later alloc. (short) | (6) #later alloc. (long) | (7) #later alloc. (short) | (8) #later alloc. (long) |
|---|---|---|---|---|---|---|---|---|
| H | -0.003 | 0.003 | -0.006 | 0.003 | 0.055 | 0.367 | -0.005 | 0.339 |
|   | (0.013) | (0.002) | (0.013) | (0.002) | (0.340) | (0.278) | (0.335) | (0.275) |
|   | [0.784] | [0.189] | [0.653] | [0.221] | [0.872] | [0.186] | [0.988] | [0.219] |
| B | -0.002 | 0.005** | -0.002 | 0.004** | 0.070 | 0.550** | 0.061 | 0.540** |
|   | (0.013) | (0.002) | (0.012) | (0.002) | (0.333) | (0.269) | (0.330) | (0.268) |
|   | [0.852] | [0.039] | [0.847] | [0.041] | [0.832] | [0.041] | [0.854] | [0.045] |
| Constant | 0.719*** | 0.924*** | 0.774*** | 0.930*** | 1.705* | 0.792 | 3.240*** | 1.553* |
|   | (0.036) | (0.006) | (0.045) | (0.007) | (0.961) | (0.770) | (1.169) | (0.916) |
|   | [0.000] | [0.000] | [0.000] | [0.000] | [0.076] | [0.304] | [0.006] | [0.090] |
| Observations | 717 | 716 | 717 | 716 | 721 | 721 | 721 | 721 |
| R-squared | 0.289 | 0.338 | 0.305 | 0.344 | 0.315 | 0.344 | 0.331 | 0.350 |
| Enum. FE | Yes | Yes | Yes | Yes | Yes | Yes | Yes | Yes |
| Controls | No | No | Yes | Yes | No | No | Yes | Yes |
| MCG+ | 0.742 | 0.929 | 0.742 | 0.929 | 2.475 | 1.383 | 2.475 | 1.383 |

Note: OLS estimates. Robust standard errors in parentheses and p-values in brackets. ***p < 0.01, **p < 0.05, *p < 0.1. Controls are age, gender, sufficient income (equal to 1 if they have enough money to feed the family in last 7 days), and education. + MCG refers to the Mean for the Control Group (R treatment). Subjects making inconsistent choices are excluded from the analysis of beta-delta.

This evidence suggests that, in the field, hypothetical (vs. real) incentives increase the variance of responses in TD tasks. However, this is true for long-term but not short-term discounting. The increase in long-term discounting SD is about 21%, which means that in order to obtain identical 95% confidence intervals for the estimations, the sample in *H* must be almost 50% larger than in *R*.

According to these data, the results from the field can be summarized as follows:

**R3 (field: H vs. R)**: *Using hypothetical vs. real payoffs does not generate different discount factors (β-δ model) or different numbers of later allocations in terms of averages. However, hypothetical decisions display larger variance of responses over the long-term (but not the short-term).*

Table 7: **Variance ratio test for the outcome variables (Study II)**

|  | (1) Beta | (2) Delta | (3) #later alloc. (short) | (4) #later alloc. (long) |
|---|---|---|---|---|
| *i) Standard deviation by treatment* | | | | |
| SD(R) | 0.145 | 0.023 | 3.818 | 2.888 |
| SD(H) | 0.139 | 0.028 | 3.873 | 3.480 |
| SD(B) | 0.143 | 0.029 | 3.999 | 3.689 |
| *ii) R vs H* | | | | |
| SD(R)/SD(H) | 1.043 | 0.821 | 0.986 | 0.829 |
| P (ratio < 1) | 0.764 | 0.002*** | 0.413 | 0.002*** |
| *iii) R vs B* | | | | |
| SD(R)/SD(B) | 1.014 | 0.793 | 0.956 | 0.783 |
| P (ratio < 1) | 0.589 | 0.000*** | 0.241 | 0.000*** |

Note: The null hypothesis is that the ratio between the standard deviation of the variable in the R group and the standard deviation in the H (or B) group is equal to 1.

As we found in the lab, TD data gathered from hypothetical MPLs are essentially not different from those obtained with real payments. Hence the core of **R1** from the lab is replicated in the field with higher statistical power.

### 5.3.2. Are BRIS and real choices different (Q2)?

Now we compare BRIS results with those from real payments. Columns 1-2 in Table 6 show that the *B* dummy yields a positive and statistically significant effect on delta ($p = 0.04$), while it is not significant for beta ($p = 0.85$). Adding controls does not change the picture (columns 3-4). Similarly, *B* is significantly positive for later allocations in the long-term block (with and without controls, columns 6 and 8, $p < 0.05$), but non-significant for allocations in the short-term block.

Regarding the variance of responses, Panel *iii)* in Table 7 shows that the ratio of the SD of the outcome variables between *R* and *B* is not significantly different from 1 for beta and short-term later allocations ($p > 0.24$). However, the variance in both delta and long-term later allocations is significantly higher in *B* compared to *R* ($p < 0.01$). The increase in SD is about 27%, meaning that to get identical 95% confidence intervals for the estimations, the sample in *B* must be about 60% larger than in *R*. Importantly, note that *B* does not yield smaller SD than *H* for any of the outcome variables, but even slightly larger.

These results suggest that subjects facing the BRIS device exhibit higher long-term patience and make noisier long-term choices than those being paid for sure. Therefore, we conclude the following:

**R4 (field: B vs. R)**: *Paying one out of ten vs. all subjects generates higher long-term patience (δ and number of later allocations in the long-term block). There is no effect for short-term discounting. In addition, BRIS increases the variance of responses over the long- but not the short-term.*

## 6 Robustness

In this section, we conduct a series of robustness checks to stress test the results obtained earlier (R1 to R4). First, we explore equivalence tests and then we move to alternative specifications.

### 6.1 Equivalence tests

Results R1 and R3 suggest that the use of hypothetical vs. real payments does not lead to different choices. On the other hand, R2 and R4 suggest that BRIS may generate some biases, although somewhat weak, which, moreover, do not coincide between the two studies.

However, regarding the non-significant estimates, it should be noted that the fact that *p*-values are larger than *alpha* (i.e. 0.05, or 0.10 for marginal significance) does not certify the absence of effect (Wagenmakers, 2007). They only tell us that we cannot reject the hypothesis that the effect is zero. To reject the hypothesis that the effect is different from zero, that is, to conclude that the *true effect size is exactly zero* we would need a huge sample size (Lakens, 2018).

One reasonable alternative is to ask whether the observed effect is large enough to be deemed worthwhile. This technique is called *equivalence testing* (ET; Lakens, 2017; Wellek, 2010) and is based on testing whether the observed effect falls within or outside an equivalence interval, defined by two predetermined bounds: the lower ($-\gamma_L$) and the upper bound ($\gamma_U$).

To test for equivalence, a two one-sided test (TOST) approach is applied in which two composite null hypotheses are tested: $H_{01} \rightarrow \gamma \leq -\gamma_L$ and $H_{02} \rightarrow \gamma \geq \gamma_U$. When both null hypotheses are rejected, we can conclude that $-\gamma_L < \gamma < \gamma_U$ or, in other words, that the observed effect falls within the equivalence bounds and it is close enough to zero to be practically equivalent (Lakens, 2017).

The challenge of this procedure is to objectively define the lower and upper bounds of the equivalence interval. In this paper, we follow Lakens (2017) and set these bounds based on benchmarks for a small size effect[6]. Specifically, we use the standardized difference value of Cohen's $d = 0.3SD$.

Following Lakens et al. (2018) we analyze not only equivalence (ET) but also the null hypothesis significance test (NHST). According to these two tests, there are four possible outcomes in the analysis. The observed effect can be (see Table S1 in the supplementary materials):

- both statistically indistinguishable from zero and statistically equivalent ($-\gamma_L < \gamma < \gamma_U$) – this is labeled as Equivalence (*E*);
- statistically different from zero and not statistically equivalent (Relevant Difference, *RD*);
- statistically different from zero but statistically equivalent (Trivial Difference, *TD*);
- neither statistically different from zero nor statistically equivalent (Undetermined, *U*).

Increasing (reducing) the Cohen's *d* used to determine equivalence would lead to a greater (smaller) probability of obtaining an Equivalence result, or E.

Figure 2 shows the coefficients obtained from regressing each outcome on *H* (triangles), *B* (squares) and a set of controls (i.e. models in columns 7-8), divided by the standard deviation of each outcome in treatment $R$ in order to be expressed in Cohen's $d$ units. The figure also displays their 90% CI, and the upper ($d = 0.3SD$) and lower bound ($d = -0.3SD$) of the equivalence interval (red vertical lines). To conclude *Equivalence*, the 90% CI line should cross the zero-effect line (grey vertical line) but not the vertical red dashed lines[7].

---

[6] Although the use of these benchmarks is typically recommended as a last resort (Lakens, 2017; Lakens et al., 2018), we stick to these bounds in the absence of a clear recommendation. We have not been able to find in the literature similar experimental designs (and estimated effects) as ours.

Tables S4 and S5 in the supplementary materials provide a detailed analysis of the results of the TOST procedure for the lab, field and online experiment.

### 6.1.1 Is *H* equivalent to *R*?

In Figure 2, the top-left graphs in panels A and B analyze if $H$ is equivalent to $R$ for beta in the lab and field experiment, respectively. In both studies, the 90% CI for $H$ crosses the zero-effect line and does not include any of the equivalence bounds. This result suggests that measuring beta with hypothetical or real incentives yields the very same results, that is, both measures are *equivalent*.

The bottom-left graphs repeat the analysis for the number of later allocations in the short-term block task. In the lab (panel A), the analysis suggests that $H$ vs. $R$ is not statistically different from zero, but they are not equivalent because the 90% CI for $H$ reaches the equivalence upper bound slightly (*Undetermined*). We would just need to increase the upper bound to about $d = 0.35SD$ to obtain that both measures are equivalent. In the case of the field (panel B), the 90% CI for $H$ crosses the zero-effect line and excludes both equivalence bounds, suggesting that $H$ and $R$ are *equivalent* regarding the number of later allocations in the short-term block.

Regarding delta (top-right graphs in Figure 2), the results are nearly identical as for short-term later allocations: in the lab (panel A) we get *undetermined* results, and would need to increase the upper bound to about $d = 0.35SD$ to obtain equivalence between $H$ and $R$; in the field (panel B), both measures are *equivalent* for delta.

---

[7] We follow Lakens et al. (2018) and use the 90% CI because in this way two one-sided tests are performed with an $\alpha = 5\%$ each

Figure 2: **Equivalence tests of the results.**

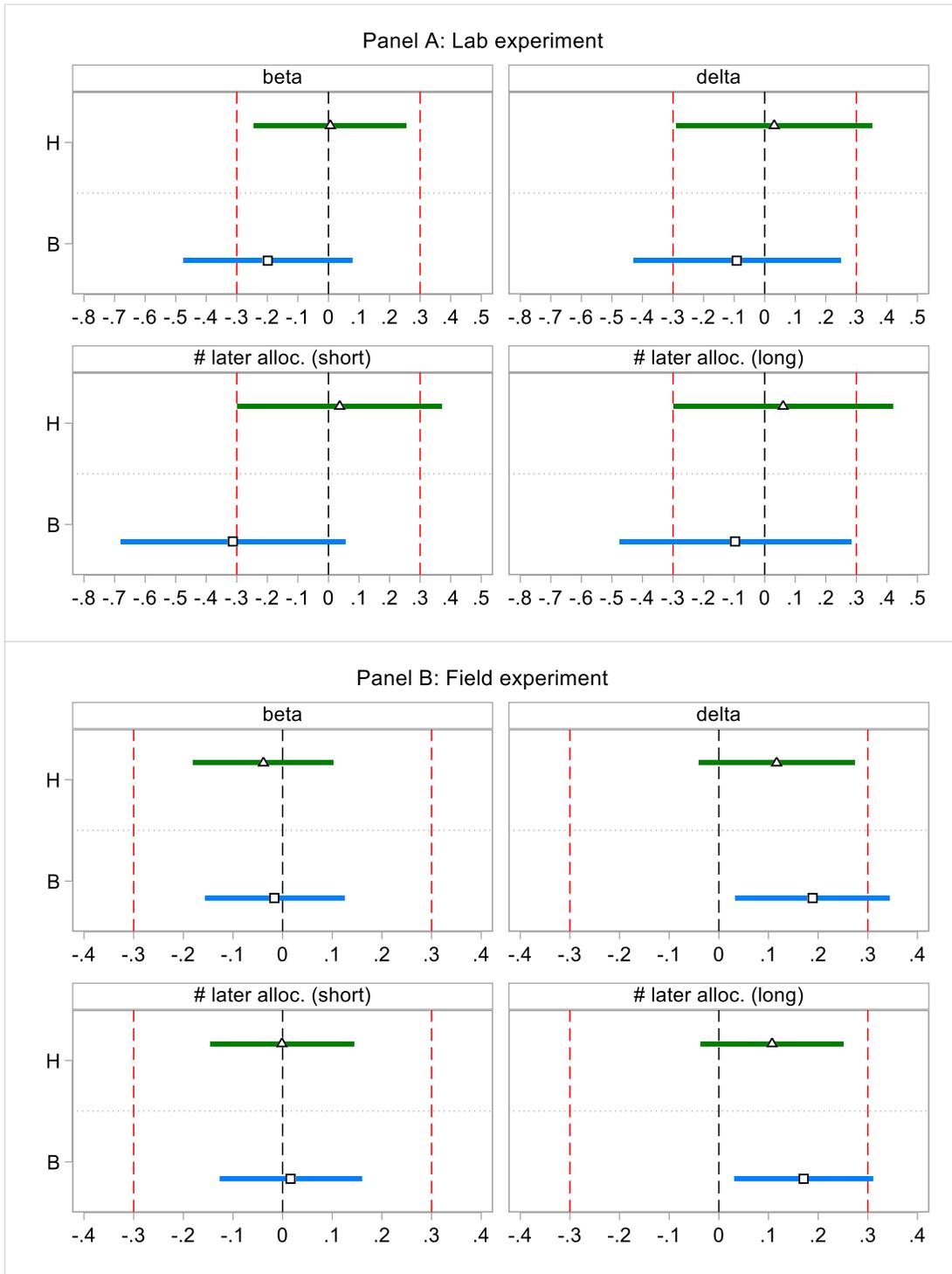

Note: Figure 2 plots the estimated standardized coefficients for *H* and *B* (vs. *R*) with their 90% CI, and equivalence bounds set to γL =-0.3SD and γU = 0.3SD (vertical red dashed line). All the models are estimated using the same controls as in models in columns 7-8 of the regression table of each study.

Finally, the bottom-right graphs in Figure 2 provide the results for the number of later allocations in the long-term block. In the lab (panel A), the 90% CI for *H* includes both zero and the equivalence upper bound. This result suggests that neither *H* vs. *R* is statistically different from zero nor they are equivalent (*Undetermined*). In this case, to get equivalence we would need to increase the upper bound to about $d = 0.40SD$. Regarding the field experiment, the 90% CI for *H* includes the zero line and excludes both equivalence bounds, thus indicating *equivalence* between *H* and *R* for the number of long-term later allocations.

All in all, the results from the lab suggest that *H* and *R* yield nearly equivalent measures for short-term discounting (especially beta, since for the number of later allocations in the short-term block we get undetermined results), whereas for long-term discounting decisions equivalence remains undetermined. In the field, we find that *H* and *R* measures are equivalent for all the outcome variables. Note that in no case we find either Trivial (TD) or Relevant Differences (RD) between *H* and *R*, and that the undetermined cases from the lab are very close to the boundaries of equivalence.

**6.1.2 Is B equivalent to R?**

This section repeats the same analysis to test whether *B* and *R* yield equivalent TD measures.

The top-left graphs in Figure 2 show the results for beta. In the case of the lab (panel A), equivalence is *undetermined*, that is, neither *B* vs. *R* is statistically different from zero nor they are statistically equivalent since the equivalence lower bound is included in the 90% CI for *B*. To get equivalence, we would need a non-trivial increase in the equivalence lower bound (up to about $d = -0.50SD$). However, in the field (panel B), the 90% CI for *B* includes zero and excludes both equivalence bounds, suggesting that both measures are *equivalent* for beta.

The bottom-left graphs in Figure 2 provide the results for the number of later allocations in the short-term block. In the case of the lab, equivalence between *B* and *R* is *undetermined* because the 90% CI for *B* includes both zero and the equivalence lower bound. Note that to get equivalence, we would need to more than double the size of the lower bound to about $d = -0.70SD$. For the field, the 90% CI for *B* includes the zero effect and excludes both equivalence bounds, suggesting that *B* and *R* are *equivalent* to measure the number of short-term later allocations.

The top-right graphs show the results for the delta discount factor. In the lab, the 90% CI lines for *B* cross both the zero-effect line and the equivalence lower bound, suggesting that equivalence for *B* vs. *R* is *undetermined*. To get equivalence, we would need an increase in the equivalence lower bound up to about $d = -0.45SD$. In the case of the field, the 90% CI for *B* excludes the zero-effect line and includes the equivalence upper bound, suggesting that *B* and *R* are not equivalent but there exists a *Relevant Difference* (RD) between both measures for delta: compared to *R*, in the field, *B* yields higher estimates for delta.

Finally, for the number of later allocations in the long-term block (bottom-right graphs) the results from the lab suggest that neither *B* vs. *R* is statistically different from zero nor they are equivalent (*Undetermined*). Again, equivalence would require the lower bound to increase to about $d = -0.50SD$. However, in the field the result shows that *B* and *R* measures are not equivalent but there exists a *Relevant Difference* (RD) between them because the 90% CI for B does not include the zero effect but includes the equivalence upper bound: in the field, *B* yields higher estimates for the number of long-term later allocations compared to *R*.

Taken together, the evidence from the lab (Study I) suggests that equivalence between *B* and *R* is largely undetermined for all the four measures considered, that is, *B* and *R* are neither equivalent nor different. On the other hand, the results in the field (Study II) are mixed: for beta and the number of later allocations in the short-term block *B* and *R* are equivalent, while for long-term discounting (delta and number of long-term later allocations) there exists a relevant difference between both measures. Note that in the field *B* yields higher estimates for long-term patience, whereas in the lab, if anything, *B* yields lower estimates for patience, especially over short-term decisions.

## 6.2 Alternative specifications: Interval and negative-binomial regressions

To account for the fact that $\beta$ and $\delta$ were actually measured in intervals, and thus all observations are either right- or left-censored or both, we re-estimate the regressions using the interval regression technique (Int-Reg; see e.g. Harrison et al., 2002). In the previous analyses, following Espín et al. (2019a), $\beta$ and $\delta$ were set to the upper value of the interval corresponding to the decision in which the participant switched from the sooner to the later option. The Int-Reg method allows us to avoid choosing an arbitrary value within the interval (e.g. the lower, upper, or central value) for calculations since these values are estimated for each interval in the regressions. On the other hand, to avoid concerns about the use of OLS, we also estimate the regressions for the number of later allocations using a negative binomial technique for count data. The regression analyses can be found in Tables S2 and S3 (supplementary materials) for the lab and the field experiment, respectively.

Figure 3 compares the estimates from OLS (triangles) for both $H$ and $B$ (vs. $R$) with those obtained using interval regressions, for $\beta$-$\delta$ (left side, squares), and negative binomial regressions, for the number of later allocations (right side, squares).

Panel A displays the results for the lab (Study I): the OLS estimates are virtually identical to the alternative estimates both for $\beta$-$\delta$ and for the number of short- and long-term later allocations. Both estimation methods yield non-significant coefficients for $H$ and $B$ on all the four TD measures according to their 95% CIs (we use 95% CIs because here we deal with two-tailed tests).

Panel B refers to the field experiment (Study II). Here we observe some small discrepancies in that the alternative specifications seem to report in general slightly stronger differences than the OLS method. However, scaling differences need to be considered: note that none of the non-significant coefficients becomes significant with the alternative method, and vice versa.

All in all, these results suggest that our findings from Studies I and II are robust to alternative regression methods.

Figure 3: **Estimated coefficients plots from different specifications.**

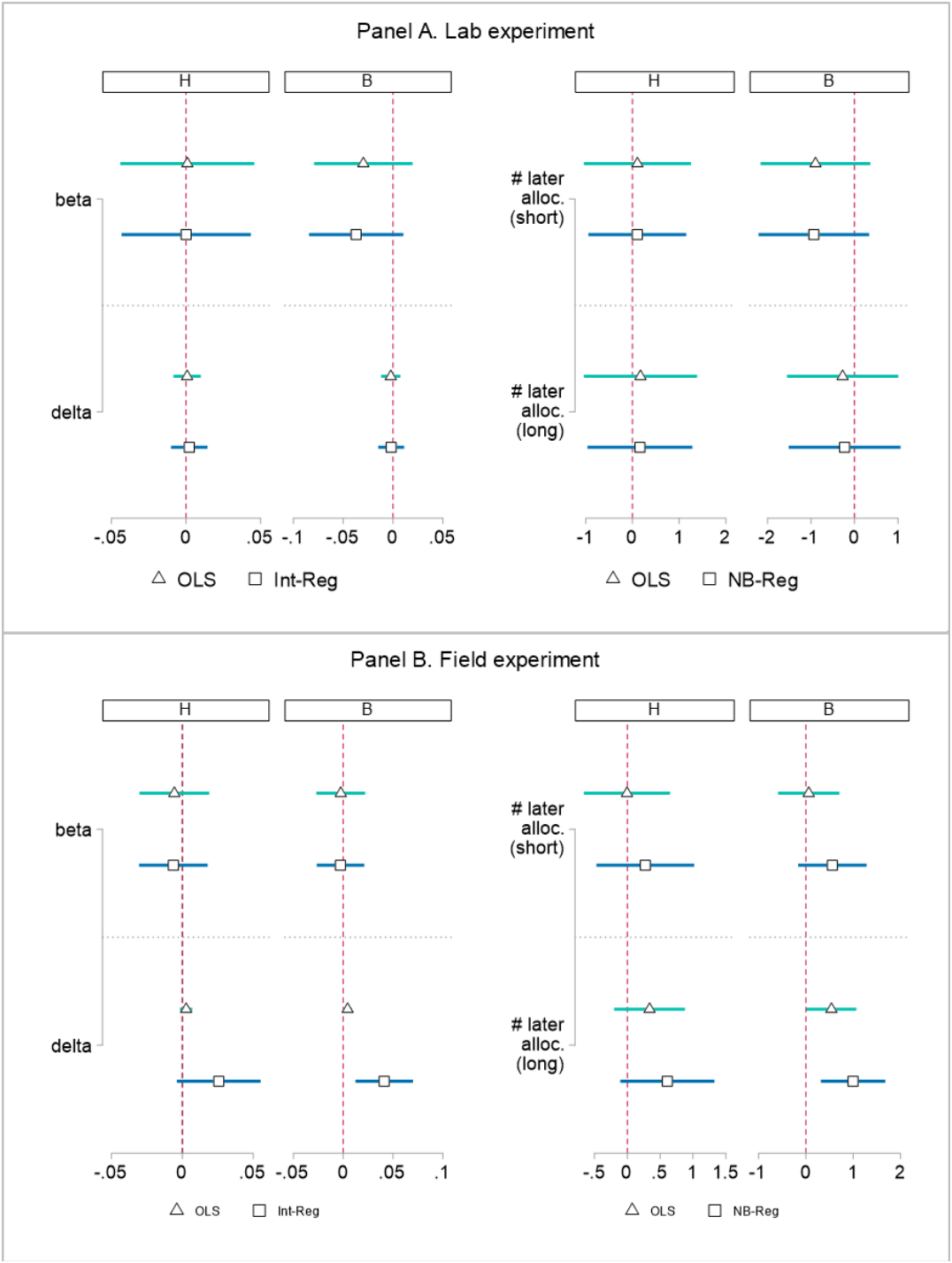

Note: Figure 3 plots the estimated coefficients for H and B, and their 95% CI from different model specifications. Int-Reg refers to interval regression and NB-Reg to negative binomial regression. All the models use the same controls as before.

# 7 Study III: The online experiment

In the previous sections, we have shown that hypothetical payoffs provide basically the same information as real incentives in TD tasks. However, the BRIS mechanism may generate different results. In particular, data gathered using BRIS yielded larger (and noisier) estimates for long-term patience in Study II.

This section is devoted to an online experiment which allows us to answer a number of important questions[8]. The purpose of this last study is twofold. First, subsection 7.3.1 compares hypothetical payoffs and BRIS in a sample of 633 subjects. Here we test whether BRIS exhibits again any particular difference. Hence, we are not asking whether hypothetical payments are informative of real ones, but whether BRIS is performing similarly to hypothetical payments. This is complemented with a comparison between *H* and *B* for Studies I and II.

Second, subsection 7.3.2 explores in detail how certain design features, which are common in large-scale experiments and surveys, affect discounting using hypothetical payments. Thus, we want to know more about the performance of hypothetical incentives, given that they seem to be a valid alternative to real ones in TD elicitation. In particular, we study the impact of within-task order (i.e. either the short-term or the long-term block first), and the possible contamination arising from the existence of previous (paid) tasks.

## 7.1 Implementation and sample

Study III has a different design than Studies I and II. To answer the aforementioned questions, we implemented a 2x2x2x2 between-subjects design. Subjects were randomly assigned to each condition.

The first arm refers to the use of BRIS vs. hypothetical payments. The other three arms refer to the within-task order, the position of the task, and the use of other paid (vs. hypothetical) tasks before the TD task. The entire sample consists of 637

---

[8] Note that online studies are becoming increasingly popular and recent advances suggest that, in fact, online data are reliable (Horton et al., 2011; Rand, 2012; Arechar et al., 2018).

subjects and 23 made inconsistent choices. The distribution by treatments is as follows:

*Hypothetical vs BRIS*: The first arms refers to the use of BRIS ($B$, $n = 315$) or Hypothetical ($H$, $n = 315$) payment schemes.

*Within-task order*: Here we explore whether deciding first either for the short- or the long-term block makes any difference in hypothetical TD. Particularly, we randomly assigned the order of the two blocks: short → long, or long → short (with 332 and 305 observations, respectively).

*Position of the task*: This arm refers to the order of the task within the entire experiment. We combined experiments with strategic interaction (games) with TD. While in Study I and II the TD task was set to be always in the first and third place, respectively, in Study III we used two sequences: TD → games, or games → TD (with 357 and 280 observations respectively).

*Previous paid tasks:* Finally, we test the effect of having other tasks which involve real money within the same experimental setup on the elicitation of hypothetical time preferences. Particularly, we randomly assigned subjects to play all other tasks (strategic games) with either hypothetical or BRIS incentives. Hence the two arms are: the other tasks within the experiment are BRIS vs. hypothetical (with 314 and 323 observations respectively). Actually, since having other (paid) tasks after TD elicitation should not affect the latter because subjects did not learn the payment method before facing each specific block of tasks (i.e. either the games or the TD), we specifically test the interaction between the variables "other tasks are BRIS vs. hypothetical" and "other tasks are before vs. after TD elicitation".

To conduct the experiment, we designed an online platform. The experiment was run between July and August of 2014. Ibercivis Foundation, based in Zaragoza, helped us to disseminate the experiment through its network of collaborators to recruit participants. They used Twitter and other social media to invite people to participate. No other restriction than having an email address and being at least 18 years old was imposed.

As in previous studies, we followed a number of procedures to ensure trust and reduce issues related to payment-uncertainty and transaction costs. These procedures were clearly explained in the instructions. Participants selected for real payments (1 out of 10 among those under BRIS) were notified the same day by email. Identically to Studies I and II we randomly selected one out of the 20 MPL decisions to compute final payoffs. We used Amazon gift cards – with specified dates – to pay winners.

Participants faced the same MPL task as in the previous studies with monetary amounts equivalent to a one-day minimum wage (initial amount = 30 euros, see Table 2). Participants who were selected to be paid earned 32.5 euros on average. We also elicited self-reported risk aversion based on three hypothetical questions.

Participants were on average 39 years old, 49% had completed university education, 23% were unemployed, and had an average income of 1,031 euros (see Table 1).

### 7.2 Ethics

All participants signed an informed consent and the data were anonymized in accordance with the Spanish Law on Personal Data Protection 15/1999.

### 7.3 Results

This section is organized in two blocks. First, we compare hypothetical data and BRIS. We first focus on the online data and then check the lab and field data for comparison reasons. Second, we study the sensitivity of hypothetical data to different settings.

### 7.3.1 Are hypothetical and BRIS choices different?

To answer the question whether $H$ and $B$ produce the same outcomes we use the entire sample of the online experiment. Table 8 provides the results of the OLS regressions. The main explanatory variable is $B$ which captures whether data arising from BRIS are different from those gathered with hypothetical incentives. Models in columns 3-4 and 7-8 include controls for age, gender, education level, income, risk preferences and *treatment* effects (see 7.1 for details). For comparability, at the bottom of Table 8 we show the results from the same regression analysis using data from Study I and II (*B-lab* and *B-field*, respectively).

Table 8: **Results from the online experiment (Study III)**

|  | (1) | (2) | (3) | (4) | (5) #later alloc (short) | (6) #later alloc (long) | (7) #later alloc (short) | (8) #later alloc (long) |
| --- | --- | --- | --- | --- | --- | --- | --- | --- |
| VARIABLES | beta | delta | beta | delta |  |  |  |  |
| *B* | 0.014 | 0.005** | 0.012 | 0.005** | 0.370 | 0.637** | 0.291 | 0.565** |
|  | (0.009) | (0.002) | (0.009) | (0.002) | (0.233) | (0.277) | (0.234) | (0.276) |
|  | [0.111] | [0.019] | [0.183] | [0.038] | [0.114] | [0.022] | [0.213] | [0.041] |
| Constant | 0.895*** | 0.951*** | 0.864*** | 0.940*** | 6.923*** | 4.316*** | 6.055*** | 3.007*** |
|  | (0.007) | (0.002) | (0.026) | (0.006) | (0.195) | (0.232) | (0.686) | (0.770) |
|  | [0.000] | [0.000] | [0.000] | [0.000] | [0.000] | [0.000] | [0.000] | [0.000] |
| Observations | 610 | 624 | 606 | 620 | 633 | 637 | 627 | 631 |
| R-squared | 0.014 | 0.014 | 0.030 | 0.055 | 0.007 | 0.014 | 0.030 | 0.053 |
| Controls | No | No | Yes | Yes | No | No | Yes | Yes |
| MCG+ | 0.786 | 0.955 | 0.786 | 0.955 | 6.860 | 4.835 | 6.860 | 4.835 |
| *B-lab* | -0.034 | -0.003 | -0.031 | -0.003 | -1.100** | -0.450 | -1.010* | -0.412 |
|  | (0.021) | (0.004) | (0.022) | (0.004) | (0.528) | (0.562) | (0.539) | (0.577) |
|  | [0.110] | [0.493] | [0.160] | [0.532] | [0.041] | [0.426] | [0.065] | [0.478] |
| *B-field* | 0.001 | 0.002 | 0.003 | 0.003 | 0.014 | 0.278 | 0.079 | 0.312 |
|  | (0.012) | (0.002) | (0.012) | (0.002) | (0.327) | (0.297) | (0.328) | (0.301) |
|  | [0.965] | [0.340] | [0.812] | [0.296] | [0.966] | [0.349] | [0.809] | [0.302] |

Note: OLS estimates. Robust standard errors in parentheses and p-values in brackets. ***p < 0.01, **p < 0.05, *p < 0.1. Controls are age, gender, income, education, risk preferences, and treatment. + MCG refers to the Mean for the Control Group (H treatment). Subjects making inconsistent choices are excluded from the analysis of beta-delta.

From the online data, can see that the dummy *B* is never significant on beta or short-term later allocations ($p > 0.11$; columns 1, 3, 5, 7), which suggests that the use of BRIS vs. hypothetical payments does not affect short-term patience elicitation. Models in columns 2, 4, 6, and 8 report a significant effect when it comes to long-term discounting: *B* increases both delta ($p < 0.05$; columns 2, 4) and the number of long-term later allocations ($p < 0.05$; columns 6, 8). In short, this implies that BRIS inflates patience over long-term outcomes compared to hypothetical payments in Study III.

On the other hand, the effect of *B* on short-term later allocations is negative and significant for the *lab* data (*B-lab*; column 5, $p = 0.04$), although it becomes marginally significant when controls are included (column 7, $p = 0.07$). Regarding beta, delta, and long-term later allocations, no significant effect is found in the *lab* ($p > 0.11$). No single effect yields significance for the *field* data (*B-field*; $p > 0.29$). Thus, in contrast to the online data, in the lab and the field, there seems to be no robust or systematic difference between *B* and *H*.

Finally, to compare the variance of responses, Table 9 shows the variance ratio test for each outcome variable. It can be seen in panel *i)* that *H* and *B* display no different variance for any outcome in the *online* data ($p > 0.43$). Similarly applies to the *lab* and *field* data (panels *ii* and *iii*; $p > 0.68$). In the latter cases, while largely insignificant, even the direction of the effect is consistently against our initial hypothesis (i.e. noisier data in *H*) as *H* always displays smaller SD. These results suggest that *H* does not increase the variance of responses compared to *B*.

We therefore conclude:

**R5 (Online: *B* vs. *H*)**: *Paying one out of ten subjects vs. none of them generates higher long-term patience ($\delta$ and number of later allocations in the long-term block). There is no robust effect for short-term discounting. In addition, BRIS does not affect the variance of responses compared to hypothetical payments.*

Even if the significant effect observed in the online sample aligns well with the result from the field study that BRIS increases patience (compared to *R*) when long-term discounting is considered, the comparison between *B* and *H* is not significant in the field. In sum, taken together, our data indicate that BRIS payments exhibit an erratic behavior and can therefore generate undesired biases.

Table 9: **Variance ratio test for the outcome variables (Study III)**

|  | (1) Beta | (2) Delta | (3) #later alloc. (short) | (4) #later alloc. (long) |
|---|---|---|---|---|
| *i) Online* | | | | |
| $SD(H)$ | 0.110 | 0.028 | 2.951 | 3.489 |
| $SD(B)$ | 0.107 | 0.028 | 2.922 | 3.498 |
| $SD(B)/SD(H)$ | 0.973 | 1.000 | 0.990 | 1.003 |
| $P\,(ratio < 1)$ | 0.635 | 0.572 | 0.430 | 0.516 |
| *ii) Lab* | | | | |
| $SD(H)$ | 0.089 | 0.018 | 2.205 | 2.388 |
| $SD(B)$ | 0.095 | 0.020 | 2.511 | 2.636 |
| $SD(B)/SD(H)$ | 1.067 | 1.111 | 1.139 | 1.104 |
| $P\,(ratio < 1)$ | 0.764 | 0.778 | 0.790 | 0.730 |
| *iii) Field* | | | | |
| $SD(H)$ | 0.139 | 0.028 | 3.873 | 3.48 |
| $SD(B)$ | 0.143 | 0.029 | 3.999 | 3.689 |
| $SD(B)/SD(H)$ | 1.029 | 1.036 | 1.033 | 1.060 |
| $P\,(ratio < 1)$ | 0.691 | 0.833 | 0.684 | 0.816 |

Note: The null hypothesis is that the ratio between the standard deviation of the variable in the B group and the standard deviation in the H group is smaller than 1.

### 7.3.2 Sensitivity of hypothetical payoffs to different settings

Along this subsection we study how sensitive hypothetical time preferences are to the within-task order between short- and long-term blocks, to the presence of other games before the TD elicitation, and to whether these are paid (BRIS). From Study III, we therefore only include the observations from the *H* treatment ($n = 307$).

Table 10 shows the results of the stress test. Models 1, 3, 5, and 7 test the main effects of the three dummies that represent the three treatments (i.e. *Games first*, *Long first*, and *Paid games*) on $\beta$, $\delta$, short- and long-term later allocations, respectively. On the other hand, models 2, 4, 6, and 8 add the interactions between the three treatment variables. All the models control for age, gender, education level and household income. *Education* and *female* have a significant impact on long-term and beta

($p<0.01$).

Table 10: **Results from the online experiment (Study III): Stress test to H**

|  | (1) | (2) | (3) | (4) | (5) # later alloc. (short) | (6) # later alloc. (short) | (7) # later alloc. (long) | (8) # later alloc. (long) |
|---|---|---|---|---|---|---|---|---|
|  | beta | beta | delta | delta |  |  |  |  |
| Games first | 0.036*** | 0.037* | 0.005 | 0.003 | 1.096*** | 1.068** | 0.695* | 0.449 |
|  | (0.012) | (0.021) | (0.003) | (0.005) | (0.322) | (0.522) | (0.388) | (0.636) |
|  | [0.004] | [0.071] | [0.123] | [0.564] | [0.001] | [0.042] | [0.074] | [0.481] |
| Long first | -0.022* | -0.025 | 0.003 | 0.003 | -0.428 | -0.720 | 0.345 | 0.231 |
|  | (0.013) | (0.023) | (0.003) | (0.005) | (0.326) | (0.596) | (0.384) | (0.652) |
|  | [0.082] | [0.275] | [0.354] | [0.614] | [0.191] | [0.228] | [0.370] | [0.723] |
| Paid games | -0.003 | 0.000 | 0.006** | 0.008 | -0.023 | 0.043 | 0.741* | 0.918 |
|  | (0.012) | (0.023) | (0.003) | (0.005) | (0.324) | (0.587) | (0.386) | (0.668) |
|  | [0.827] | [0.996] | [0.048] | [0.149] | [0.945] | [0.942] | [0.056] | [0.170] |
| Games first*Long first |  | 0.005 |  | 0.004 |  | 0.407 |  | 0.558 |
|  |  | (0.025) |  | (0.006) |  | (0.645) |  | (0.771) |
|  |  | [0.851] |  | [0.541] |  | [0.529] |  | [0.470] |
| Games first*Paid games |  | -0.008 |  | 0.000 |  | -0.339 |  | -0.052 |
|  |  | (0.025) |  | (0.006) |  | (0.640) |  | (0.763) |
|  |  | [0.762] |  | [0.981] |  | [0.597] |  | [0.945] |
| Long first*Paid games |  | 0.002 |  | -0.003 |  | 0.209 |  | -0.300 |
|  |  | (0.026) |  | (0.006) |  | (0.673) |  | (0.792) |
|  |  | [0.948] |  | [0.615] |  | [0.756] |  | [0.705] |
| Constant | 0.884*** | 0.884*** | 0.936*** | 0.936*** | 6.344*** | 6.350*** | 2.509** | 2.547** |
|  | (0.038) | (0.039) | (0.009) | (0.009) | (0.966) | (0.982) | (1.104) | (1.119) |
|  | [0.000] | [0.000] | [0.000] | [0.000] | [0.000] | [0.000] | [0.024] | [0.024] |
| Observations | 307 | 307 | 314 | 314 | 318 | 318 | 319 | 319 |
| R-squared | 0.046 | 0.047 | 0.055 | 0.057 | 0.057 | 0.059 | 0.056 | 0.058 |
| Controls | Yes | Yes | Yes | Yes | Yes | Yes | Yes | Yes |
| MCG+ | 0.881 | 0.881 | 0.952 | 0.952 | 6.757 | 6.757 | 4.553 | 4.553 |

Notes: OLS estimates. Robust standard errors in parentheses and p-values in brackets. ***p < 0.01, **p < 0.05, *p < 0.1. + MCG refers to the Mean for the Control Group (i.e. the three treatment dummies = 0).

The elicitation of both $\beta$ and the number of short-term later allocations is sensitive to *Games first* ($p < 0.01$; columns 1, 5). If other games are played before the TD task, subjects show higher level of short-term patience, according to both measures. Since the interaction *Games first*Paid games* is not significant (indeed, none of the interactions tested is ever significant; $p > 0.50$), the positive effect of *Games first* on short-term patience holds regardless of whether the games are paid or not (see columns 2 and 6). In addition, the non-significant interaction between *Games first* and *Long first* suggests that within-task order does not moderate the effect of *Games first*. Also, the sequence long→short (vs. short→long), captured by *Long first* is marginally associated to a lower beta ($p = 0.08$; column 1). Nevertheless, the remaining regressions suggest that this is not a robust effect.

The elicitation of $\delta$, on the other hand, is robust both to other games being played before and to different within-task orders, while is apparently sensitive to the use of monetary incentives in other tasks: *Paid games* yields a positive and significant effect ($p = 0.05$; column 3). The effect is similar but marginally significant for long-term later allocations ($p = 0.06$; column 7). Yet, given that the interaction *Games first\*Paid games* is never significant (see columns 4 and 8), this should be considered a spurious result. Since subjects could not know ex-ante whether the games would be paid (BRIS) or hypothetical, we should expect the interaction to be positive and significant, indicating that the observed positive effect of *Paid games* only exists when the games are played first but *not* when TD is first. We instead find a similar effect in both conditions. Finally, as for beta and short-term later allocations, playing other games first (paid or not) marginally increases the number of long-term later allocations ($p = 0.07$; column 7).

Therefore, we can conclude that:

**R6:** *Hypothetical time preferences are robust to different within-task orders (long/short) and to whether other tasks are incentivized. However patience, especially in the short-term, is larger if the TD task comes after other experimental tasks.*

In sum, hypothetical time preferences are fairly robust to several settings. The fact that placing the TD elicitation after other tasks affects the results is of particular interest for large-scale experiments and surveys in which a number of different tasks are typically introduced in the same questionnaire.

# 8 General discussion

This paper performs a systematic study of the impact of different incentive schemes in the elicitation of time preferences using MPLs. We cover lab, field and online experiments with very different subject pools.

Our results from lab and field experiments suggest that non-incentivized (hypothetical) decisions provide similar measures as incentivized decisions in the elicitation of time preferences using MPLs. We also observe that the variability of responses in short-term TD measures (beta, or present bias, and number of later allocations in the short-term block) does not differ between hypothetical and real scenarios. However, in the field sample we find about 21% higher variance in hypothetical vs. real responses for long-term TD measures (delta and the number of later allocations in the long-term block). Taking all together, our data suggest that subjects in the lab and in the field display comparable temporal preferences when elicited using hypothetical vs. real rewards. These results are robust to different estimation procedures and, in a non-negligible number of cases, we can even conclude that both methods elicit *equivalent* TD measures.

These findings are in line with the literature from the lab (Johnson and Bickel, 2002; Madden et al., 2003, 2004; Lawyer et al., 2011; Matusiewicz et al., 2013), and with the scarce results available from the field (Ubfal, 2016; Harrison et al., 2002).

There are two important implications. First, our study demonstrates that existing hypothetical MPL tasks, often used to gather TD data in the field, are indeed informative of individual time preferences. At least, they are essentially not different from those elicited with real incentives. Our findings therefore indicate that hypothetical time preferences are a valid proxy for incentivized ones and that therefore payments may be dispensed with. Eliminating real payoffs also reduces other problems: children can more easily participate in experiments, subjects do not need to release private information, transaction costs, inflation rates, etc. This is good news for field studies that may include this sort of TD tasks with a minimal impact on their budgets. All in all, having a hypothetical but reliable measurement of time preferences is relevant for several reasons. Time discounting can be measured on large samples or even on the entire population.

We also find that BRIS payment schemes may lead to different measurements compared to incentivized decisions. Although our field (and, indirectly, online) data indicate that this might be true especially for long-term TD measures, our results also suggest that the impact of BRIS might be erratic, that is, the bias does not always arise in the same direction. Consequently, our main recommendation is to either pay to all participants or not pay anyone.

We also showed that hypothetical TD measures are robust to several settings that are often used in surveys and large-scale experiments. However, our analysis suggests that playing other games before the TD elicitation, regardless of whether these are also hypothetical or not, may lead subjects to display higher patience, especially on short-term TD measures. Future research should explore this result in more detail.

There are, at least, two important limitations. First, although our studies do not cover only typical experimental subjects (i.e. self-selected students; see Exadaktylos et al. 2013) still we have missed a key share of the population: kids and adolescents. Eliciting time preferences in kids is critical for policy evaluation (Levitt et al., 2016; Giné et al., 2017). Understanding children's time preferences is essential to have a more comprehensive understanding of their choices in domains such as education, sexual behavior, drug abuse, etc. As mentioned, in any case, our study might be especially important for this type of research since the use of real money with non-adult samples is particularly complicated.

Second, our study focuses on MPLs. While there is an intense debate over whether Convex Time Budgets (CTBs) work better than MPLs, the latter have been used more extensively. A precise analysis of the impact of hypothetical decisions on CTBs in necessary. An exception is the recent work by Brañas-Garza and Prissé (2020) who compare hypothetical, real, and BRIS payments using continuous MPL – a procedure in between MPL and CTB – and find that hypothetical decisions yield similar results as incentivized ones. This suggests that the current findings can be extended to other elicitation tasks, but more research is needed.

# Supplementary Information
Paid and hypothetical time preferences are the same: Lab, field and online evidence

## A. Equivalence Testing

In sections 4.1.3 and 5.1.3 we find that some estimated coefficients of *H* and *B* are not significant. The null hypothesis significance testing (NHST) cannot support the conclusion of absence of effect (Wagenmakers, 2007). Statistical equivalence testing (ET) is more appropriate for testing the absence of an effect. There the null hypothesis is that two measures are different by at least as much as an equivalence interval defined by some chosen level of tolerance. The alternative hypothesis is that the measurements are statistically equivalent (Wellek, 2010). So, the acceptance of the alternative hypothesis gives strong support to the interpretation that there is no effect of the independent variable (payment treatment) on the outcomes variable. That is, both measurements are equivalent in terms of the outcome variable.

As we mention in Section 6.1, we define the tolerance level to be equal to Cohen's d=0.3SD. That is, for each outcome variable, we compute the upper (lower) bound as the coefficient plus (less) 30% of the standard deviation of the outcome in the Real group. This defines the equivalence interval for each estimated coefficient.

To test for equivalence, a two one-sided test (TOST) approach is applied in which two composite null hypotheses are tested: $H_{01} \rightarrow \gamma \leq -\gamma L$ and $H_{02} \rightarrow \gamma \geq \gamma U$. When both null hypotheses are rejected, we can conclude that $-\gamma L < \gamma < \gamma U$ or, in other words, that the observed effect falls within the equivalence bounds and it is close enough to zero to be practically equivalent (Lakens, 2017).

The challenge of this procedure is to objectively define the lower and upper bounds of the equivalence interval. In this paper, we follow Lakens (2017) and set these bounds based on benchmarks for a small size effect. Specifically, we use the standardized difference value of Cohen's d = 0.3SD.

Taking into account the results from NHST and ET, we can obtain four possible conclusions according to Tryon and Lewis (2008). Table S1 summarizes the four possible results.

Table S1: **Possible conclusions in equivalence testing.**

| NHST | ET | Conclusion |
|---|---|---|
| Not Reject $H_0$ | Reject $H_{01}$ and $H_{02}$ | Equivalence (E) |
| Reject $H_0$ | Not Reject $H_{01}$ and $H_{02}$ | Relevant difference (RD) |
| Reject $H_0$ | Reject $H_{01}$ and $H_{02}$ | Trivial Difference (TD) |
| Not Reject $H_0$ | Not Reject $H_{01}$ and $H_{02}$ | Undetermined (U) |

Table S2 provides the equivalence test results for the different outcomes using the TOST approach[9]. It can be appreciated from Table S4 that $H$ and $R$ measures are equivalent only for beta. However, the p values of the rests of variables are closed enough to $\alpha = 0.1$, suggesting that a very small increase in the equivalence interval (e.g., from 0.3SD to 0.35SD) will lead to equivalence conclusions. Different is the case of $B$ where the p values are far from $\alpha = 0.1$. In this case the conclusion is that we do not have enough statistical power to do the test.

---

[9] To perform the equivalence test, we use the tostregress command from Stata 16 developed by A. Dinno (2017). URL:https://www.alexisdinno.com/stata/tost.html

Table S2: **Equivalence test results for the lab (Study I).**

| Outcome | | (1) beta | (2) $p(H_{01})$ | (3) $p(H_{02})$ | (4) Change | (5) Conclusion |
|---|---|---|---|---|---|---|
| beta | H | 0.001 | 0.094 | 0.081 | 0.031 | E |
| | B | -0.030 | 0.482 | 0.008 | 0.031 | U |
| delta | H | 0.001 | 0.060 | 0.109 | 0.006 | U |
| | B | -0.002 | 0.192 | 0.040 | 0.006 | U |
| # later alloc. (short) | H | 0.106 | 0.060 | 0.116 | 0.800 | U |
| | B | -0.897 | 0.561 | 0.004 | 0.800 | U |
| # later alloc. (long) | H | 0.171 | 0.054 | 0.145 | 0.822 | U |
| | B | -0.274 | 0.199 | 0.046 | 0.822 | U |

Note: Column 1 shows the estimated coefficient for H and B from the OLS regressions with controls and their significance level from the NHST (***$p < 0.01$, **$p < 0.05$, *$p < 0.1$). Columns 2 and 3 show the p-values from the two one-side-test for the two-null hypothesis in the equivalence test (ET). Columns 4 show the level used to perform the TOST (30% of each outcome's SD); and column 5 shows the conclusion considering both test NHST and ET. The baseline is the group making decisions with real money (R).

Table S3 shows the equivalence test results for the field. Hypothetical ($H$) and real ($R$) decisions yield equivalent measures of patient in the field. However, $B$ and $R$ are equivalent only for beta and the number of later allocations in the short-term block, while both measures are different in long-term decisions. Again, the p-values of $B$ are close to $\alpha = 0.1$ suggesting that an small increase in the equivalence bounds will lead to conclude that both measures are trivially different. This implies that the effect of $B$ is close to a small size effect.

Table S3: **Equivalence test results for the field (Study II).**

| Outcome | | (1) beta | (2) p($H_{01}$) | (3) p($H_{02}$) | (4) Eq. Level | (5) Conclusion |
|---|---|---|---|---|---|---|
| beta | H | -0.006 | 0.001 | 0.000 | 0.044 | E |
| | B | -0.002 | 0.001 | 0.000 | 0.044 | E |
| delta | H | 0.003 | 0.000 | 0.028 | 0.007 | E |
| | B | 0.004** | 0.000 | 0.131 | 0.007 | RD |
| # later alloc. (short) | H | -0.005 | 0.003 | 0.004 | 0.900 | E |
| | B | 0.061 | 0.002 | 0.006 | 0.900 | E |
| # later alloc. (long) | H | 0.339 | 0.000 | 0.030 | 0.860 | E |
| | B | 0.540** | 0.000 | 0.117 | 0.860 | RD |

Note: Column 1 shows the estimated coefficient for H and B from the OLS regressions with controls and their significance level from the NHST (***$p < 0.01$, **$p < 0.05$, *$p < 0.1$). Columns 2 and 3 show the p-values from the two one-side-test for the two-null hypothesis in the equivalence test (ET). Columns 4 show the level used to perform the TOST (30% of each outcome's SD); and column 5 shows the conclusion considering both test NHST and ET. The baseline is the group making decisions with real money (R).

## B. Interval regressions

As robustness checks, we run interval regressions of beta and delta on the different payments treatments dummies. These variables are measured in intervals, and thus that all observations are right and left censored. So as robustness check, we re-estimate the regressions using interval regression techniques. We also run a negative binomial model for the number of later allocations in the short and long run task. The results for the lab are shown in Table S4. The coefficients of the treatments variables are very similar than those estimated in Table 4: *H* is never significant while *B* is marginally significant for beta and the number of later allocations in the short-run.

Table S5 provides the results of the interval and negative binomial regressions for the field experiment. The coefficients of the treatments variables are very similar than those estimated in Table 6: *H* is never significant while *B* is significant for delta and the number of later allocations in the long-run. This suggest BRIS mechanism payments have an impact on longer delay decisions.

Table S4: **Interval regressions for the lab (Study 1)**

| | (1) | (2) | (3) | (4) | (5) | (6) | (7) | (8) |
|---|---|---|---|---|---|---|---|---|
| | | | | | # later alloc. (short) | # later alloc. (short) | # later alloc. (long) | # later alloc. (long) |
| | beta | beta | delta | delta | | | | |
| $H$ | -0.003 | 0.002 | 0.001 | 0.002 | -0.000 | 0.095 | 0.104 | 0.160 |
| | (0.022) | (0.006) | (0.022) | (0.006) | (0.507) | (0.543) | (0.534) | (0.573) |
| | [0.898] | [0.795] | [0.967] | [0.720] | [1.000] | [0.861] | [0.845] | [0.780] |
| $B$ | -0.037* | -0.003 | -0.030 | -0.002 | -1.144* | -0.363 | -0.934 | -0.228 |
| | (0.022) | (0.006) | (0.025) | (0.007) | (0.602) | (0.620) | (0.649) | (0.655) |
| | [0.097] | [0.646] | [0.222] | [0.789] | [0.057] | [0.558] | [0.150] | [0.728] |
| Constant | 0.843*** | 0.929*** | 0.839*** | 0.933*** | | | | |
| | (0.016) | (0.005) | (0.059) | (0.019) | | | | |
| | [0.000] | [0.000] | [0.000] | [0.000] | | | | |
| Observations | 116 | 120 | 114 | 118 | 120 | 120 | 118 | 118 |
| Controls | No | No | Yes | Yes | No | No | Yes | Yes |
| MCG+ | 0.839 | 0.937 | 0.839 | 0.937 | 5.601 | 2.701 | 5.601 | 2.701 |

Note: Robust standard errors in parentheses and p-values in brackets. ***p < 0.01, **p < 0.05, *p < 0.1.

Table S5: **Interval regressions for the field (Study II)**

| | (1) | | (2) | (3) | (4) | (5) | (6) | (7) | (8) |
|---|---|---|---|---|---|---|---|---|---|
| | | | | | | # later alloc. (short) | # later alloc. (short) | # later alloc. (long) | # later alloc. (long) |
| | beta | | beta | delta | delta | | | | |
| $H$ | -0.004 | 0.019 | -0.006 | 0.015 | 0.272 | 0.673* | 0.256 | 0.614 | -0.004 |
| | (0.013) | (0.013) | (0.012) | (0.013) | (0.378) | (0.355) | (0.381) | (0.557) | (0.013) |
| | [0.741] | [0.147] | [0.607] | [0.252] | [0.472] | [0.058] | [0.502] | [0.270] | [0.741] |
| $B$ | -0.003 | 0.028** | -0.003 | 0.025** | 0.476 | 0.972*** | 0.522 | 1.011* | -0.003 |
| | (0.012) | (0.013) | (0.012) | (0.013) | (0.368) | (0.338) | (0.372) | (0.573) | (0.012) |
| | [0.819] | [0.028] | [0.813] | [0.048] | [0.197] | [0.004] | [0.161] | [0.078] | [0.819] |
| Constant | 0.723*** | 0.813*** | 0.779*** | 0.850*** | | | | | 0.723*** |
| | (0.036) | (0.035) | (0.045) | (0.041) | | | | | (0.036) |
| | [0.000] | [0.000] | [0.000] | [0.000] | | | | | [0.000] |
| Observations | 717 | 716 | 717 | 716 | 721 | 721 | 721 | 721 | 717 |
| Controls | Yes | Yes | Yes | Yes | Yes | Yes | Yes | | Yes |
| MCG+ | No | No | Yes | Yes | No | No | Yes | Yes | No |

Note: Robust standard errors in parentheses and p-values in brackets. ***p < 0.01, **p < 0.05, * p <0.1.